\documentclass[pdflatex,sn-mathphys-num]{sn-jnl}

\usepackage{graphicx}%
\usepackage{multirow}%
\usepackage{amsmath,amssymb,amsfonts}%
\usepackage{amsthm}%
\usepackage{mathrsfs}%
\usepackage[title]{appendix}%
\usepackage{xcolor}%
\usepackage{textcomp}%
\usepackage{manyfoot}%
\usepackage{booktabs}%
\usepackage{algorithm}%
\usepackage{algorithmicx}%
\usepackage{algpseudocode}%
\usepackage{listings}%
\usepackage[utf8]{inputenc}
\theoremstyle{thmstyleone}%
\theoremstyle{thmstyletwo}%

\theoremstyle{thmstylethree}%

\newcommand{\ep}{\varepsilon}
\newcommand{\Nmax}{N_{2,\mathrm{max}}}
\newcommand{\Neq}{N_{2,\mathrm{eq}}}

\begin{document}

\title[Article Title]{Nonlinearity Reversal in Epsilon-Near-Zero Indium Tin Oxide Driven by Few-Cycle Light Pulse}


\author[1]{\fnm{Mustafa Goksu} \sur{Ozlu}}
\equalcont{These authors contributed equally to this work.}

\author[1]{\fnm{Colton} \sur{Fruhling}}
\equalcont{These authors contributed equally to this work.}
\author[1]{\fnm{Ian} \sur{Hoffman}}
\author[1]{\fnm{Jae-Ik}\sur{Choi}}
\author[2]{\fnm{Marcello}\sur{Ferrera}}
\author[1]{\fnm{Alexandra}\sur{Boltasseva}}
\author*[1]{\fnm{Vladimir M.}\sur{Shalaev}}\email{shalaev@purdue.edu}

\affil[1]{\orgdiv{Birck Nanotechnology Center}, \orgname{Purdue University}, \orgaddress{\city{West Lafayette}, \state{IN}, \country{USA}}}

\affil[2]{\orgdiv{Institute of Photonics and Quantum Sciences}, \orgname{Heriot-Watt University}, \orgaddress{\city{Edinburgh}, \country{UK}}}


\abstract{\small Recent breakthrough studies of nonlinearities at extreme pump intensities ($\sim$1 $\text{TW/cm}^2$) in transparent conducting oxides (TCOs) have rewritten our understanding of the dynamics in these materials. However, exploring TCO dynamics beyond these intensities is prohibited by the damage threshold of the material. In this work, we overcome this problem by using a few-cycle pump laser pulse (sub-8\,fs) to maximize the intensity while keeping the optical fluence below the damage threshold. We observe a reversal in the optical response trend starting at optical pump laser intensities of $\sim$5 $\text{TW/cm}^2$ similar to Segal et al\cite{segalComplex2025}. At the highest pump pulse intensities, we obtain a complete change in the sign of the modulation for both transmission and reflection, producing a full-cycle oscillation of the refractive index modulation within 300\,fs. The amplitude of the sign reversal scales quadratically with the intensity. We therefore propose a simple two-photon absorption (TPA) model to explain the observed behaviour. The TPA, which is normally forbidden by the Pauli blocking, is enabled here by intraband excitations from the lower to the upper non-equilibrium states of the conduction band (CB). Such excitations vacate the states at the bottom of the CB, lifting up the blocking and thus making interband TPA possible. The model is in good agreement with experimental results, capturing the essential trends in the observed data and revealing the dynamics of competing channels caused by the interplay between interband and intraband transitions. This intensity-controlled mechanism could be the key to unlocking new applications of TCOs for time-varying photonics such as photonic time crystals.}




\maketitle

\section{Introduction}\label{sec1}

Ultrahigh intensity optical pulses\cite{stricklandCompression1985} are unparalleled tools for revealing new light-matter interaction phenomena and enabling disruptive technologies such as high harmonic generation (HHG)\cite{lewensteinTheoryHighharmonicGeneration1994,tsurHighHarmonicGeneration2022}, molecular dynamics\cite{Attosecond2021}, attosecond physics\cite{Krausz2009}, laser plasma accelerators\cite{tajimaLaserElectronAccelerator1979}, photonic time-crystals\cite{Lyubarov2022,Park2025}, and in general time-varying photonics\cite{Galiffi2022,Engheta2023}. Transparent conducting oxides (TCOs) have become keystone materials for these types of studies because of their large and ultrafast optical response when pumped at extreme intensities\cite{Alam2016,capretti2015,Kinsey2015}. This is particularly true in the spectral range where the real part of the dielectric permittivity $\varepsilon$ is near zero (ENZ). In TCOs, the ENZ crossing is governed by free electrons in the conduction band described by the Drude model. One such TCO is indium tin oxide (ITO), whose electron density places the ENZ in the technologically relevant telecom frequency band. The ENZ regime in TCOs has been used to enhance many optical phenomena such as harmonic generation\cite{caprettiEnhancedThirdharmonicGeneration2015,capretti2015}, optical time reversal\cite{Vezzoli2018}, all-optical switching\cite{cleary2018,feigenbaum2010,fan2023,Bohn2021}, adiabatic frequency shifting\cite{Khurgin2020a,Zhou2020,Liu2021,bohnSpatiotemporalRefractionLight2021a}, switchable gain\cite{Pendry2021,Jaffray2024} and coherent absorption\cite{galiffi2026}.

The large optical response in TCOs arises from real electronic transitions driven by optical absorption of a pump photon\cite{Khurgin2020}. If the photon energy is above the bandgap (interband), electrons are added to the conduction band and the plasma frequency increases. Conversely, if the photon energy is below the bandgap (intraband), electrons are excited to higher energies within the conduction band, where the electron effective mass is larger. This has the opposite effect of decreasing the plasma frequency. Thus, the direction of optical response can be controlled by the pump-photon energy\cite{Clerici2017}. Crucially, while the the linear absorption process modifies the refractive index at all wavelengths, this change is maximized at the ENZ regime. This can be understood by examining the change in the refractive index at the ENZ, $\partial n/\partial \varepsilon = 1/\sqrt{\varepsilon}$. Clearly, this diverges when $\varepsilon =0$, leading to a breakdown in perturbation theory\cite{reshef2017,khurgin2024}. Our work here shows that as we explore towards higher pump intensity regimes, the nonlinearity in the pump absorption must also be considered.

Early demonstrations of optically driven dynamics in TCOs were conducted at intensities starting from 100s of $\text{GW/cm}^2$ and have been limited to $\sim \text{1-2 TW/cm}^2$. This limit is set by the thermal damage threshold of TCOs determined by the pulse fluence. Recent experiments have shown that novel dynamics might lay at higher intensities. For example, under ultra-intense illumination with a relatively long (225fs) pump, an unexplained rise-time of the optical response was inferred. The estimated 1-10\,fs rise-time is dramatically shorter than expected\cite{tirole2023}. One possible theory suggests that the extreme photon number in the pump triggers an avalanche process that quickly saturates the optical response\cite{pendryAvalanche2024}. In another study, a near-single-cycle pump induced not only an ultrafast rise-time, but also an ultrafast recovery\cite{Lustig2023}. Importantly, the recovery is not explained by the current two-temperature model and the results initiated multiple competing theories\cite{narimanov2026,hayranokDispersion2022,fruhling2025time}. More recently, Segal et al. reported  non-monotonic oscillations in the transmission and reflection of ITO under high-photon-flux modulation, where the transmission and the reflection changed trend \cite{segalComplex2025}, though the underlying mechanism was not uncovered.

In this work, we overcome the $~1-2\,\text{TW/cm}^2$ limit by using a near-single-cycle optical pulse (sub-8\,fs). This allows us to extend the studies to $\sim\!10\,\text{TW/cm}^2$ regime, where we discover a complete reversal of the optical response both in transmission and reflection. This builds directly on the non-monotonic behaviour first observed by Segal et. al \cite{segalComplex2025}, where at high pump intensities the change in transmission became negative while the reflection changed directions without completely switching signs. Here we extend that study and observe a complete sign reversal in both transmission and reflection channels and explain the effect through the developed model. To understand the origin, we systematically study the role of pump intensity in pump-probe experiments with ITO. We employ a sub-8\,fs pump pulse to reach unexplored high intensities while remaining at low fluences evading thermal damage. The optical response is probed near the ITO ENZ wavelength, 1225\,nm. Remarkably, above $\text{4.9 TW/cm}^2$  pump intensity, the response reverses sign. At the highest pump intensities, the maximal optical response has the same magnitude, but completely opposite sign compared to lower intensities. In addition, the sign reversal scales quadratically with the pump intensity, signalling a departure from the linear and monotonic scaling previously observed\cite{Kinsey2015,Caspani2016,Zhou2020}. To describe our results, we develop a simple phenomenological three-level model that includes two key features: two-photon absorption (TPA) and  Pauli unblocking in the non-equilibrium state. The model captures the essential non-equilibrium dynamics and reveals a full-cycle oscillation of the refractive index modulation within the first ~300\,fs. Unlike previous two-color experiments\cite{Clerici2017}, this mechanism demonstrates an intensity-controlled pathway for ultrafast time-varying media applications including optical-frequency photonic time crystals.

\section{Results}\label{sec2}

\subsection{Few-cycle Pump-probe Experiments}
\begin{figure}[h]
    \centering
    \includegraphics[width=1\linewidth]{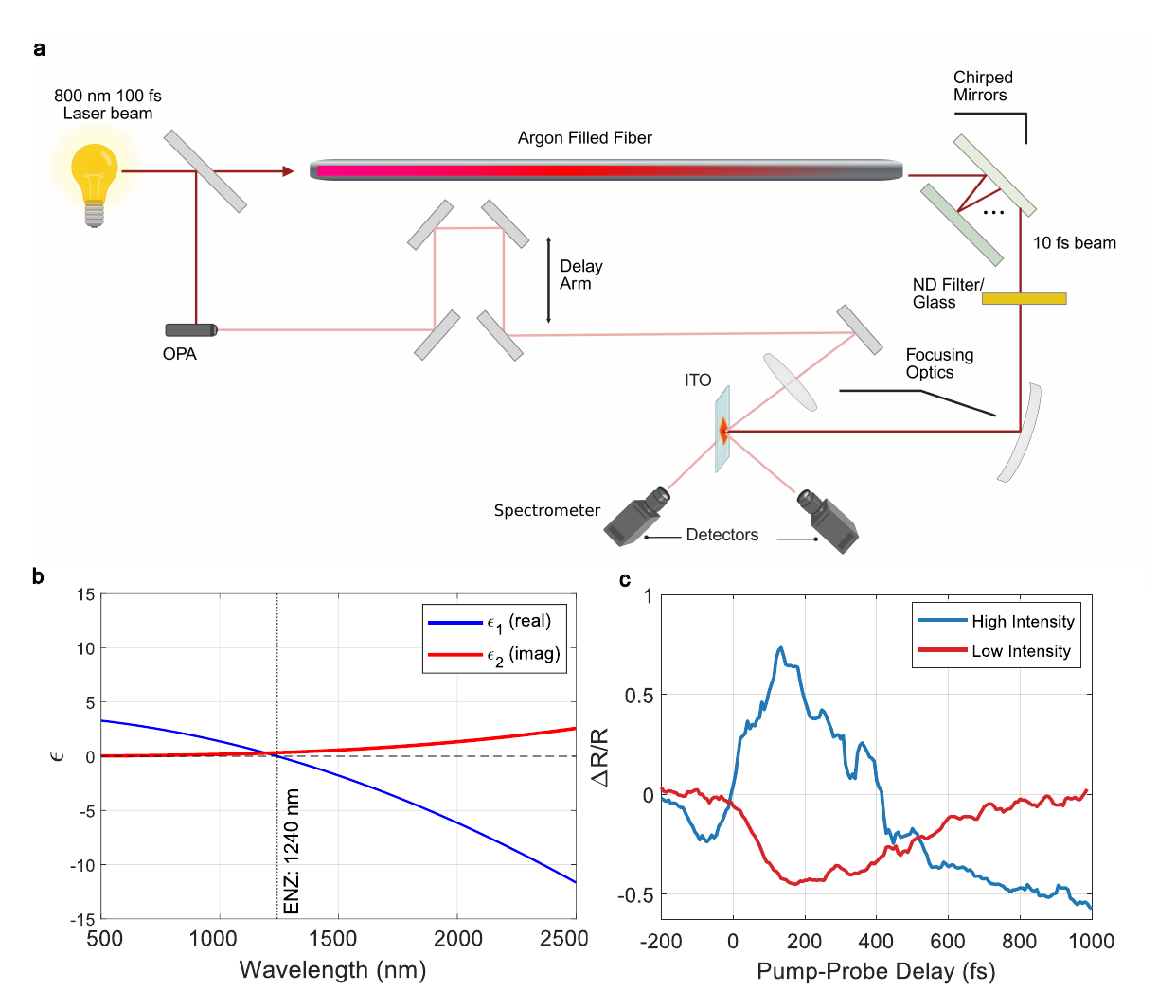}
    \caption{\textbf{Experimental setup and sample characterization.} (a) Schematic of the pump-probe setup. An 800 nm, 100 fs Ti:Sapphire output is spectrally broadened in an argon-filled hollow-core fiber and compressed to sub-8 fs by chirped mirrors. A 1225 nm probe pulse is generated by an OPA. Pump-probe delay is controlled by a mechanical delay arm on the probe path. Both beams are focused onto a 270 nm ITO film and the reflected and transmitted probe intensities are recorded simultaneously by two detectors. (b) Complex permittivity of the ITO sample extracted from spectroscopic ellipsometry and fitted with a Drude model. The real part of the permittivity crosses zero at 1240 nm.}
    \label{fig:fig1}
\end{figure}

In the conducted pump-probe experiment (Fig.\ref{fig:fig1}a) a sub-8fs\,, 800\,nm pump pulse and a 100\,fs, 1225\,nm probe pulse are focused onto an ITO sample. The use of a sub-8 fs pump enables access to very high peak fields while limiting the deposited energy and avoiding damage, thereby allowing exploration of a strongly driven, non-perturbative regime. The longer probe pulse is not intended to resolve sub-cycle dynamics, but rather to measure the effective permittivity change induced by this extreme excitation. The probe wavelength is chosen to sit at the ENZ crossing of the sample, a 270\,nm-thick ITO film deposited on a 1.1\,mm-thick glass substrate. The complex permittivity of the sample was characterized by spectroscopic ellipsometry and fitted with a Drude model (Fig.\ref{fig:fig1}b). The reflected and transmitted probe intensities and the transmitted probe spectrum are recorded as a function of the pump-probe delay. The zero delay for each scan is set by the delay where the maximum frequency shift is observed in the transmitted probe spectrum. This corresponds to the delay of maximal pump-probe temporal overlap. All dynamics discussed hereafter occur after this reference point, beyond the coherent interaction regime, and reflect genuine population and carrier dynamics rather than pump-probe interference effects\cite{dietzekAppearanceCoherentArtifact2007}. The transmitted probe spectrum as well as a control experiment with a bare glass sample are presented in the Supplementary Information.

At lower intensities, the transient change in reflection (Fig.\ref{fig:fig1}c) $\Delta R/R_0$ is negative while the change in transmission $ \Delta T/T_0$ is positive. This response is consistent with a pump-induced reduction of the plasma frequency driven by intraband transitions. Absorption of the 800 nm pump photons excites conduction-band electrons to higher-energy states in the non-parabolic band, where the larger effective mass reduces the plasma frequency\cite{Khurgin2020,Sarkar2023}. The reduction of the plasma frequency redshifts the ENZ wavelength away from the probe making the material less metallic. Thus, the reflectance decreases and the transmittance increases in accordance with the Drude picture\cite{Clerici2017}. However, at higher intensities the response is completely opposite (blue curve in Fig.\ref{fig:fig1}c. The relative change in reflection completely flips sign, signaling a breakdown in the simple description of linear absorption paired with the Drude model.

\begin{figure}[h]
    \centering
    \includegraphics[width=1\linewidth]{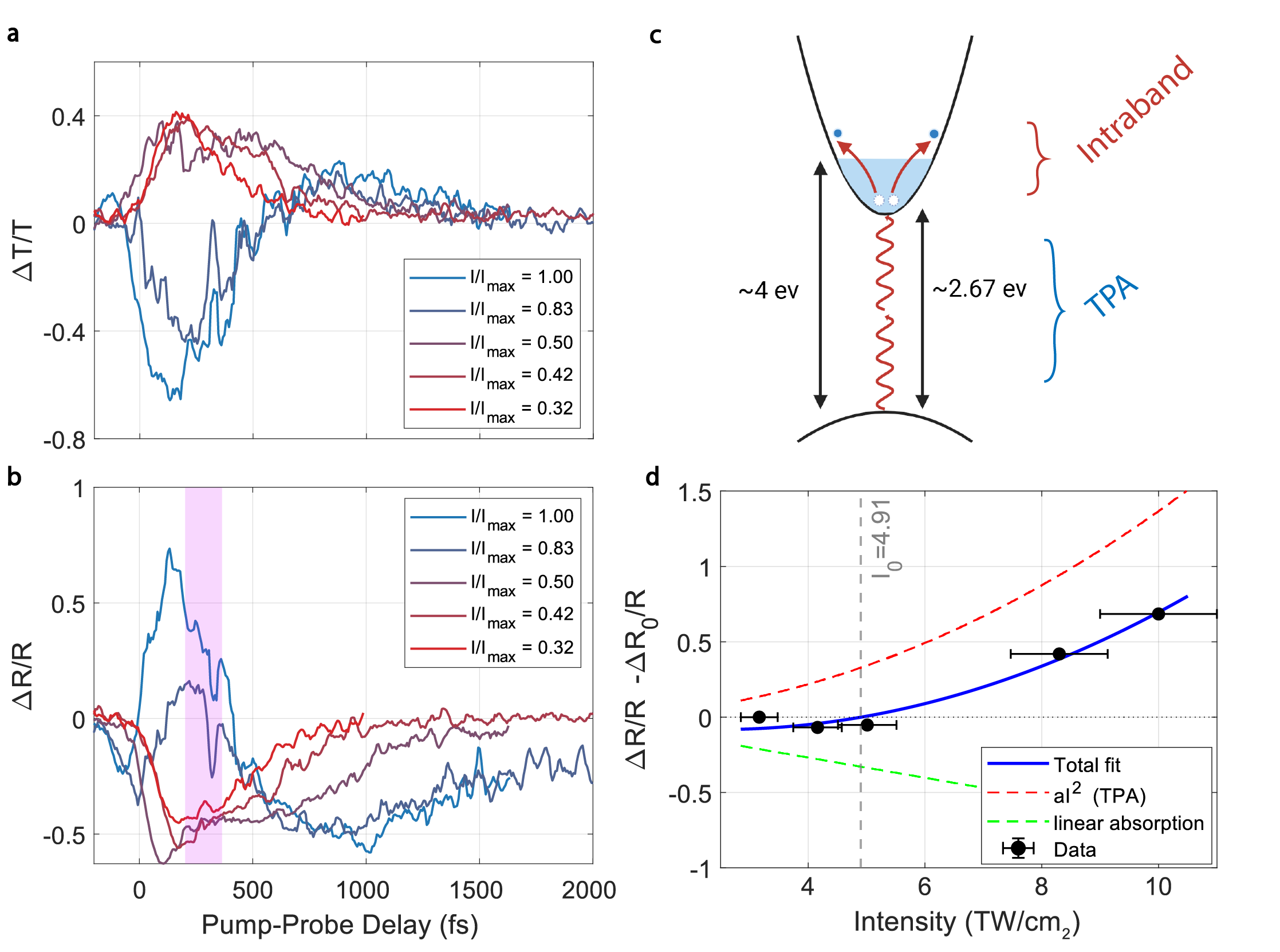}
    \caption{\textbf{Intensity-dependent pump-probe response of ITO and identification of the competing nonlinear channel.} (a,b) Normalized transient transmission $\Delta T/T_0$ and reflection $\Delta R/R_0$ as a function of pump-probe delay for five pump intensities. The pink shaded region in (b) indicates the quasi-static window used for intensity scaling analysis. (c) Schematic of the proposed mechanism: intraband pump excitation clears conduction-band states, lifting Pauli blocking and enabling two-photon interband absorption across the ~2.67 eV electronic gap. (d) Intensity scaling of the peak $\Delta R/R_0$ within the shaded window after subtracting the intraband baseline, decomposed into a linear absorption contribution (green dashed) and a quadratic two-photon absorption contribution (red dashed), with total fit (blue solid). The threshold intensity $\text{I}_0$ = 4.91 $\text{TW/cm}^2$ marks the crossover between the two regimes. }
    \label{fig:fig2}
\end{figure}

To understand this breakdown, we perform an intensity scan across the transition (Fig.\ref{fig:fig2}). As the pump intensity increases, there are several trends to notice. First, increasing the pump intensity results in a diminished maximum modulation of both $\Delta R/R_0$ and $\Delta T/T_0$ as the intraband transitions begin to saturate\cite{tirole2022}. The saturation is most clearly seen in the three lowest intensity curves in Figs. \ref{fig:fig2}a and \ref{fig:fig2}b. The second trend occurs as the intensity is increased further. We observe the gradual appearance of a dip (purple curve) and ultimately the complete sign reversal of the transient response during the first $\sim\!100$\,fs. Non-monotonic behaviour of this kind was first reported by Segal et al. \cite{segalComplex2025}, where the transmission reversed sign while the reflection showed a change in trend. Here, at higher peak intensities we observe a complete sign reversal in both channels. At highest pump intensity this reversal reaches a peak $\Delta R/R_0$ of $\sim\!0.7$, before returning to negative values and finally recovering monotonically towards zero on a  1\,ps timescale. The sign-reversing, non-monotonic dynamics at high intensity is completely inconsistent with a single-photon intraband absorption mechanism and indicates the onset of a competing nonlinear mechanisms. 

We note that this reversal feature appears approximately 150 fs after the point of the maximum spectral blueshift of the transmitted probe spectrum (see Supplementary Information), which marks the peak in the pump-probe overlap. Since this happens after the few-cycle pump left the sample, the reversal is more consistent with a real carrier redistribution involving a finite thermalization time rather than with an instantaneous virtual-transition or coherent process. Two photon absorption or linear excitation of electrons are followed by a finite-time thermalization, therefore their effects would be present even after the pump. A separate, smaller feature that can be caused by virtual transitions does appear earlier, during the pump-probe overlap, near the zero delay, at the highest intensities, but it does not exhibit the sign reversal.

To identify the intensity scaling of the competing channel and the underlying mechanism, we extract the peak $\Delta R/R_0$ within the 200–400 fs window (highlighted region in Fig.\ref{fig:fig2}b). This region corresponds to the quasi-static regime after electron thermalization, but before the onset of electron-phonon decay. The lowest pump intensity $\Delta R/R_0$ is subtracted from each curve to provide a baseline. The resulting intensity dependence (Fig.\ref{fig:fig2}d) reveals a super-linear scaling. Fitting the results to a quadratic polynomial that incorporates both single and two photon processes, we find an excellent agreement with the data. Moreover, the fit retrieves a negative linear contribution consistent with the expected effect from intraband pumping. The quadratic contribution indicative of a two-photon process is positive and grows to dominate. A gray dashed vertical line at $4.9\text{TW/cm}^2$ is plotted where the fit predicts the transition from the single-photon-dominated regime to the two-photon-dominated regime.

\bigskip

\subsection{Bandgap and Absorption Dynamics}
In degenerate ITO, the optical bandgap measured by the onset of absorption is Moss-Burstein shifted to approximately 4\,eV due to the occupation of conduction-band states up to the Fermi level\cite{Walsh2008,Walsh2008a}. However, the electronic bandedges from the valence-band maximum to the conduction-band minimum is at most 2.67\,eV in In$_{2}$O$_{3}$ \cite{Walsh2008a}, the parent compound of ITO. Tin doping shifts the Fermi level into the conduction band, but the band topology and interband gap energy are expected to remain close to those that of the parent compound \cite{Walsh2008}. Thus in equilibrium, TPA of two 800\,nm ($\sim$1.55\,eV) pump is prohibited by the Pauli blocking. 

In the non-equilibrium pumped state, intraband absorption drives electrons away from the bottom of the conduction band: filling higher level states and opening lower-level states. As the pump increases, the intraband response saturates because of the finite population of conduction-band electrons available for intraband absorption\cite{tirole2022}. This sets a natural upper bound on the plasma frequency shift, as discussed in the context of non-perturbative ENZ nonlinearity by Khurgin and Kinsey \cite{khurgin2024}. While this does slow down the growth of the response, it cannot explain the complete reversal of the optical response seen at the highest intensities.

Based on the observation of a quadratic scaling in the effect, we adopt the most simple model, TPA enabled by the relaxation of the Pauli-blockade in the non-equilibrium state. Under low pump intensities, this channel is suppressed by Pauli blocking; since the conduction band bottom is occupied and final states for interband transitions are unavailable. At elevated intensities, intraband absorption and heating redistributes conduction-band electrons to higher energy states, vacating the band bottom and lifting the Pauli blockade. This opens phase space for two-photon interband absorption, generating additional free carriers that blue-shift the plasma frequency and produce the competing positive $\Delta R/R_0$ contribution. The mechanism is illustrated schematically in Fig. \ref{fig:fig2}c. To assess whether this picture is internally consistent, we construct a three-level rate equation model describing the essential linear intraband and quadratic interband absorption, presented in the following section. It should be here noted that we are operating at extreme intensities, where higher order nonlinearities such as harmonic generation\cite{Jaffray2025un}, may also contribute. However, we restrict our model to the lowest order correction (TPA).

While this manuscript was in preparation, a related study appeared, reporting similar non-monotonic dynamics and a sign reversal of the plasma frequency in ITO under high-fluence pumping \cite{Harwood2026}. Their work independently suggests a competing interband channel as the origin of such behavior. However, the observation of a complete inversion of the response in both reflection and transmission under few cycle excitation, together with the quadratic scaling, are key aspects revealed in this work.
\bigskip

\subsection{Three-level rate equation model and permittivity dynamics}

To model the observed intensity-dependent dynamics, we employ a three-level rate equation system comprising one valence band level $|1\rangle$ and two conduction band levels: a lower level $|2\rangle$ associated with the bottom of the conduction band ($m^{*}_{2} = 0.35\,m_{0}$) and an upper level $|3\rangle$ corresponding to higher-energy conduction band states ($m^{*}_{3} = 0.50\,m_{0}$). The population dynamics are described by following rate equations:

\begin{align}
\frac{dN_{1}}{dt} &= -\beta I^{2} N_{1} f_{2}
  + \gamma_{21}\!\left(N_{2}P_{1} - \Neq P_{1,\mathrm{eq}}\right) \label{eq:N1}\\
\frac{dN_{2}}{dt} &= +\beta I^{2} N_{1} f_{2} - \gamma_{21}\!\left(N_{2}P_{1} - \Neq P_{1,\mathrm{eq}}\right) - \sigma I N_{2} + \gamma_{32} N_{3} \label{eq:N2}\\
\frac{dN_{3}}{dt} &= +\sigma I N_{2} - \gamma_{32} N_{3} \label{eq:N3} \\
P_1 &= N_{1eq} - N_1 \label{eq:P1}
\end{align}

\noindent where $f_{2} = \max\!\left(0,\, 1 - N_{2}/\Nmax\right)$ is the Pauli
blocking factor for transitions into the lower conduction band. The carrier populations are $N_{1}$, $N_{2}$, and $N_{3}$ representing the valence band, lower conduction band, and upper conduction band levels, respectively. $P_{1}$ and $P_{1,\mathrm{eq}}$ are the instantaneous and equilibrium valence band hole populations respectively. The maximum capacity of the lower conduction band level is represented by $N_{2eq}$. 

Equation \ref{eq:N1} describes the valence-band dynamics. The first term describes the removal of carriers via TPA, where $\beta$ governs the strength of the interaction. The second term describes the relaxation from the lower conduction band controlled by the decay rate $\gamma_{21}$. This term is limited by the equilibrium population $\Neq$ and the available holes. Equation \ref{eq:N2} describes the lower conduction band. The first two terms are simply the inverse of Eq. \ref{eq:N1}. The third term describes single-photon absorption to the upper conduction band $|3\rangle$, which is controlled by the parameter $\sigma$ and the linear dependence on the pump intensity $I$. The final term describes the relaxation from the upper $|3\rangle$ to the lower $|2\rangle$ conduction band via the decay rate $\gamma_{32}$. The upper conduction band is described by Eq. \ref{eq:N3} with the two terms being the inverse of the last two terms in Eq. \ref{eq:N2}. Finally, the hole population $P_1$ is related to the carriers by Eq. \ref{eq:P1}.

Physically, the rate $\gamma_{32}$ describes phonon-mediated relaxation from the upper to the lower conduction band, while $\gamma_{21}$ governs net recombination written in detailed balance form through the term $(N_{2}P_{1} - \Neq P_{1,\mathrm{eq}})$. Inclusion of $P_{1}$ and $P_{1,\mathrm{eq}}$ ensures that the system relaxes to thermal equilibrium in the absence of excitation. In the degenerate limit, appropriate for ITO, the lower conduction band is taken to be fully occupied at equilibrium ($N_{2} = \Neq$) and the equilibrium hole population is set to zero ($P_{1,\mathrm{eq}} = 0$), reflecting the Fermi level sitting well inside the conduction band.

\begin{figure}[h]
    \centering
    \includegraphics[width=1\linewidth]{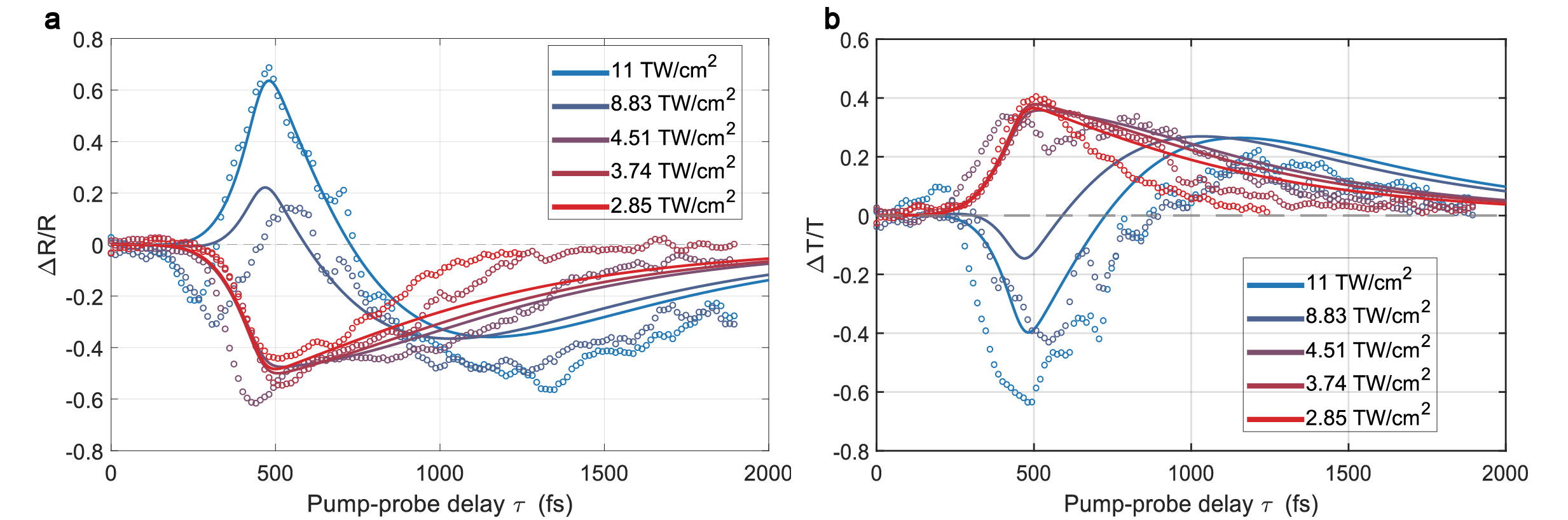}
    \caption{\textbf{Three-level rate equation model fits to the measured transient optical response.} (a) $\Delta T/T_0$ and (b) $\Delta R/R_0$ as a function of pump-probe delay for all five pump intensities (circles: data, solid lines: model). The model is fitted simultaneously to both channels across all intensities.}
    \label{fig:fig3}
\end{figure}

The time-dependent carrier populations $N_{2}(t)$ and $N_{3}(t)$ are then inserted in to a Drude permittivity model through their contributions to the plasma frequency. As the two conduction band levels correspond to states with distinct effective masses, they contribute separate Drude terms:

\begin{equation}
\ep(\omega,t) = \ep_{\infty}
  - \frac{\omega^{2}_{p_2}(t)}{\omega^{2}+i\gamma_{D_2}\omega}
  - \frac{\omega^{2}_{p_3}(t)}{\omega^{2}+i\gamma_{D_3}\omega}
\label{eq:drude}
\end{equation}

\noindent where $\ep_{\infty}$ is the high-frequency background permittivity, $\omega_{p2}$ and $\omega_{p3}$ are the time-dependent plasma frequencies associated with the lower and upper conduction band populations respectively, and $\gamma_{D2}$ and $\gamma_{D3}$ are the corresponding Drude damping rates. The plasma frequency of the lower conduction band is scaled as $\omega_{p2}(t) = \omega_{p2,\mathrm{eq}}\sqrt{N_{2}(t)/\Neq}$, preserving the equilibrium permittivity in the absence of excitation. The upper conduction band contribution $\omega_{p3}$ is initialized to zero at equilibrium and grows as carriers are promoted into level $|3\rangle$. The transient reflectance and transmittance are calculated from the time-dependent permittivity using the transfer matrix method (see Methods). The model outputs are compared directly to the measured $\Delta R/R_0$ and $\Delta T/T_0$ for all five pump intensities, as shown in Fig.~3.

The model is fitted simultaneously to the measured $\Delta R/R_0$ and $\Delta T/T_0$ for all five pump intensities using Levenberg-Marquardt least-squares optimization. Parameters with well-established values in the literature are held fixed, while the remaining parameters are allowed to vary freely. A complete list of fixed and free parameters and their fitted values is given in Table~\ref{tab:params}. To account for uncertainty in the absolute pump intensity calibration, an independent intensity scale factor per curve is allowed to vary within $\pm$10\% of the nominal value. The time evolution of the plasma frequencies and the corresponding complex permittivity from the fitted model are shown in Fig.~4.

\begin{table}[h]
\centering
\caption{Model parameters. Fixed parameters are set from independently measured or
established values. Free parameters are determined by simultaneous optimisation to
the measured $\Delta R/R$ and $\Delta T/T$ for all five intensities. CB and VB stand for conduction band and valence band respectively.}
\label{tab:params}
\small
\begin{tabular}{l l l l l}
\toprule
Parameter & Symbol & Fixed/Free & Fitted value & Units \\
\midrule
Background permittivity    & $\ep_{\infty}$         & Fixed & $3.9$                    & ---              \\
Total carrier density      & $n_\mathrm{total}$     & Fixed & $10^{27}$                & m$^{-3}$         \\
Lower CB damping           & $\gamma_{D2}$          & Fixed & $1.24\times10^{14}$      & rad\,s$^{-1}$    \\
Upper CB damping           & $\gamma_{D3}$          & Fixed & $1.24\times10^{14}$      & rad\,s$^{-1}$    \\
Lower CB effective mass    & $m^{*}_{2}$            & Fixed & $0.35$                   & $m_{0}$          \\
Equilibrium CB occupation  & $\Neq$                 & Fixed & $\Nmax$                  & ---              \\
TPA coefficient            & $\beta$                & Free  & $1.13\times10^{-12}$     & cm$^{4}$\,s      \\
Intraband cross-section    & $\sigma$               & Free  & $24.0$                   & cm$^{2}$         \\
CB3$\to$CB2 relaxation     & $\gamma_{32}$          & Free  & $1.61\times10^{12}$      & s$^{-1}$         \\
CB2$\to$VB recombination   & $\gamma_{21}$          & Free  & $1.27\times10^{13}$      & s$^{-1}$         \\
Upper CB capacity          & $N_{3,\mathrm{max}}$   & Free  & $3.00$                   & ---              \\
Upper CB effective mass    & $m^{*}_{3}$            & Free  & $0.50$                   & $m_{0}$          \\
\bottomrule
\end{tabular}
\end{table}

The model reproduces the qualitative features of the measured $\Delta R/R_0$ well across all measured intensities. The sign reversal at high intensity, the peak position, and the negative trough at intermediate delays are captured in Fig. \ref{fig:fig3}b. Although the model fits the general intensity trend well, there are certain discrepancies worth pointing out. For example, the magnitude of the change in $\Delta T/T_0$ at the two highest intensities is underestimated (Fig. \ref{fig:fig3}a). This discrepancy arises in part because the upper conduction band damping rate $\gamma_{D3}$ was held fixed during fitting. In reality, the Drude damping rate in ITO is known to evolve with the carrier temperature. Phonon and electron-electron scattering increase at elevated electron temperatures, while ionized impurity scattering decreases, making the net temperature dependence non-trivial and difficult to constrain without additional measurements \cite{wangExtendedDrudeModel2019,Gurung2025}. This is especially true for ultra-thin and transdimensional films\cite{choi2025pathwayopticalcycledynamicphotonics}. Since the transmission is more sensitive to optical losses than reflection, the error introduced by a fixed $\gamma_{D3}$ is more prominent in the transmission (Fig.\ref{fig:fig3}a). Additionally, at longer delays beyond $\sim$1200\,fs, the effects of constant relaxation rates $\gamma_{32}$ and $\gamma_{21}$ becomes apparent, as the model predicts a faster relaxation than measured.

\begin{figure}[h]
    \centering
    \includegraphics[width=1\linewidth]{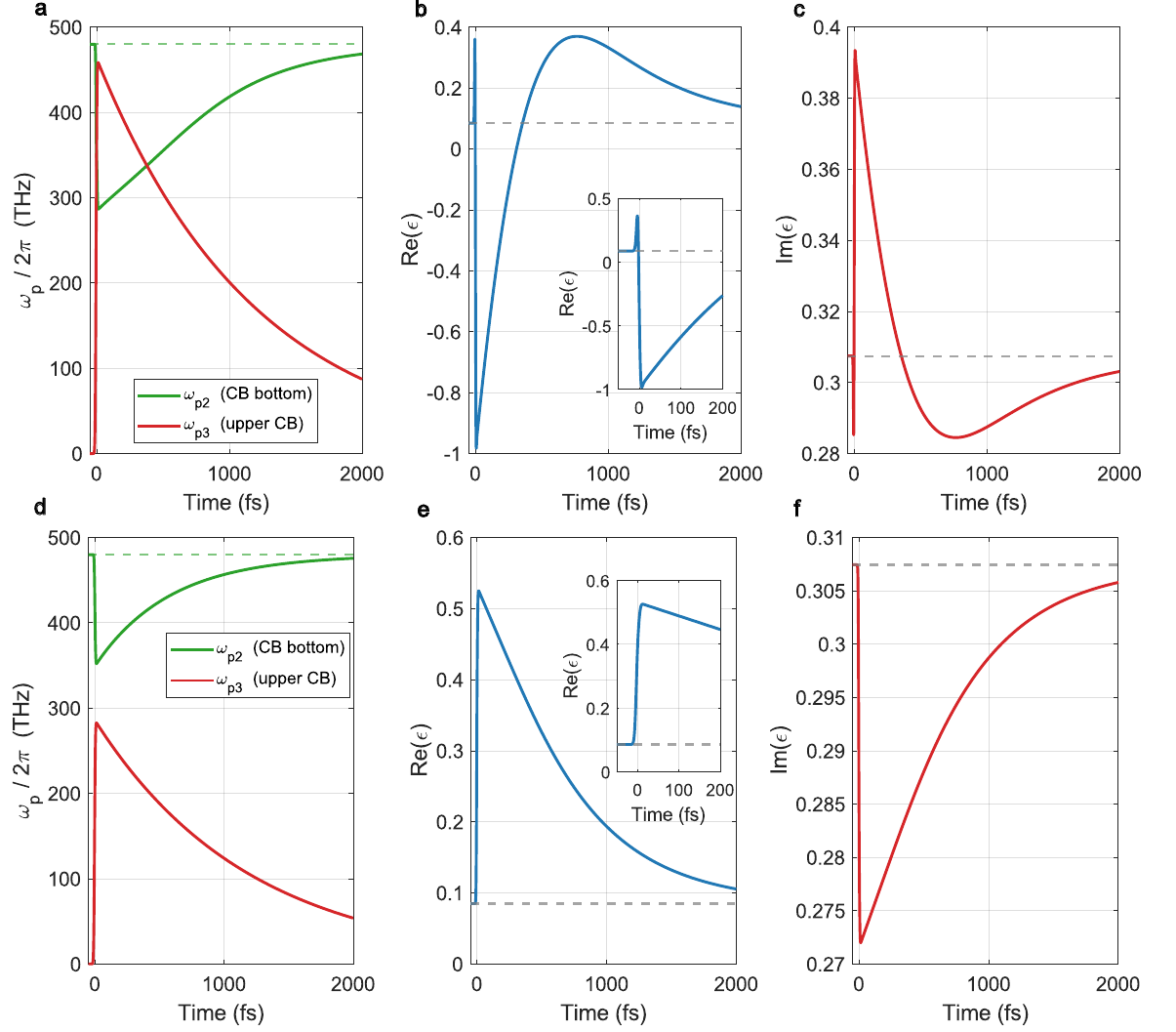}
    \caption{\textbf{Time-dependent permittivity dynamics from the fitted model.} (a,d) Time evolution of the plasma frequencies $\omega_{p2}$ (lower conduction band, green) and $\omega_{p3}$  (upper conduction band, red) at the highest and lowest pump intensities respectively. (b,e) Corresponding real part of the permittivity $\mathrm{Re} ( \varepsilon )$, with the equilibrium value indicated by dashed lines. (c,f) Imaginary part $\mathrm{Im} ( \varepsilon )$. At high intensity $\mathrm{Re} (\varepsilon)$ undergoes a complete modulation cycle, driven upward by intraband heating, sharply negative by two-photon interband absorption, and back above equilibrium during recovery. At lower intensity $\mathrm{Re}( \varepsilon )$ remains positive throughout, consistent with a purely intraband response.}
    \label{fig:fig4}
\end{figure}

Another feature present in the data is the sharp decay at early delays in $\Delta R/R_0$ visible in the two highest intensity curves. This comes from two sources not included in our model. First, the coherent pump-probe interaction is strongest at these delays and causes energy exchange between pulses\cite{dietzekAppearanceCoherentArtifact2007}. Secondly, the finite thermalization time is also not included. The model replaces the full carrier distribution with two discrete population levels, neglecting both the initial non-thermal distribution immediately following excitation and the subsequent thermalization process by which electron-electron scattering establishes a hot Fermi-Dirac distribution on tens to hundreds of femtoseconds timescales \cite{Liu2014,Sarkar2023}. This limits accuracy in the early delay regime where both effects are significant. The permittivity dynamics in Fig. \ref{fig:fig4}b shows that the initial rapid rise in $\mathrm{Re}(\ep)$ is present in the model but occurs instantaneously. Once convolved with the 100\,fs probe envelope this feature is compressed and lost. In the experiment, this rise occurs over a finite thermalization time, producing a small initial decay before the sign reversal that the model cannot capture. Despite these discrepancies, the model reproduces the essential intensity-dependent trends and provides a consistent account of the competing nonlinear channels across the full range of pump fluences.

 The model retains the essential physical ingredients of the proposed mechanism and provides a tractable framework for assessing whether Pauli-unblocking-enabled two-photon absorption is consistent with the observed intensity-dependent dynamics. The time-dependent permittivity computed from the fitted model provides direct insight into the physical mechanism of the underlying dynamics. At the highest pump intensity, shown in \ref{fig:fig4}a--c, the $\mathrm{Re}(\ep)$ begins at its equilibrium value and is first driven positive as intraband absorption reduces the plasma frequency. The onset of interband TPA then rapidly reverses this trend, generating a burst of free carriers that drive $\mathrm{Re}(\ep)$ sharply negative to nearly $-1$, pushing the material deep into the metallic regime. As the interband-generated carriers relax, $\mathrm{Re}(\ep)$ recovers back through zero and overshoots above the equilibrium value before slowly returning. The result is a complete modulation cycle of the real permittivity --- upward, sharply negative, and back above equilibrium --- completed within approximately 300\,fs. At lower intensities, shown in \ref{fig:fig4}d--f, this behavior is absent: $\mathrm{Re}(\ep)$ remains positive throughout and decays monotonically, consistent with a purely intraband response that never drives the material across the ENZ point. Importantly, this feature is also absent in the model when keeping the fluence the same but lengthening the pump pulse to 100\,fs, while keeping all other parameters fixed(See Supplementary Information). This emphasizes the importance of using a near-single-cycle pump. Finally, the ability to execute a full sub-300\,fs permittivity modulation cycle in a near-zero permittivity material is directly relevant to time-varying photonic applications, where the speed and depth of index modulation are the primary figures of merit.

\section{Conclusion}\label{sec13}

In summary, we present a systematic pump-probe study of the ultrafast nonlinear optical response of ITO under few-cycle excitation. We study the evolution of transient response with respect to the pump intensity. At low pump intensities, the response is well described by intraband absorption within the non-parabolic conduction band, producing a monotonic reduction of the plasma frequency consistent with the established picture of slow ENZ nonlinearity. At elevated pump intensities, we observe a qualitative breakdown of this behavior, culminating at the highest intensities in a complete reversal of the dominant nonlinear mechanism. We observe a sign-reversing, non-monotonic transient state that develops within $\sim$300\,fs and cannot be accounted for by intraband dynamics alone. Through a three-level rate equation model coupled to a Drude permittivity and transfer matrix optical calculation, we attribute this behavior to two-photon interband absorption enabled by pump-induced Pauli unblocking of the conduction band bottom, whose quadratic intensity scaling is directly confirmed by the experimental data. 

The few-cycle pump regime is essential to accessing the observed dynamics. At fixed fluence below the damage threshold, the sub-8\,fs pulse duration maximizes the peak intensity, pushing the system into a regime where intraband and interband channels compete and ultimately invert. The resulting sub-300\,fs full modulation cycle of the real permittivity --- driven from above ENZ to negative and back above equilibrium --- represents one of the fastest and deepest permittivity swings reported in an ENZ material. This finding is directly relevant to the ongoing pursuit of time-varying photonic media and optical-frequency photonic time crystals, where ITO has been identified as a leading candidate material precisely because of its large and fast permittivity modulation. Understanding and controlling the intensity-dependent interplay between intraband and interband nonlinear channels is a necessary step toward engineering the full permittivity trajectory on sub-cycle timescales.
\bigskip
\backmatter

\bmhead{Methods}
The pump is generated by compressing a portion of the output of a Ti:Sapphire laser through a hollow-core fiber pulse compression system and confirmed by FROG characterization. The pump intensity is varied by neutral density filters. A FROG trace for the pump can be found in the supplementary. The probe is generated by an OPA from the remainder of Ti:Sapphire laser system. Both pump and probe are s-polarized to minimize the thin-film resonance effects from exciting ENZ samples with p-polarization. The pump-probe delay is controlled via a delay arm on the probe path. Both pump and probe are focused on to the sample at incidence angles of 0$\deg$ and 20$\deg$ respectively. The pump is focused to a beam size of 750\,$\mu\text{m}$ on the sample, while the probe is focused to 50\,$\mu\text{m}$, such that the probe observes a spatially homogeneous refractive index change.  

The transfer matrix method assumes a 270\,nm ITO film on a 1.1\,mm glass substrate at 20$^{\circ}$ angle of incidence and TE polarization, convolved with the 100\,fs probe-pulse envelope to account for the finite probe duration. 

\bmhead{Supplementary information}
Additional data and simulations can be found in the Supplementary information.


\bmhead{Data Availability}
Source data and simulation code are available from the corresponding author upon request.

\bmhead{Code Availability}
MATLAB code used for rate-equation modeling and TMM fitting is available upon
request.

\bmhead{Competing Interests}
The authors declare no competing interests.

\bmhead{Funding}
This material is based upon work supported by the U. S. 
Army Research Laboratory and the U. S. Army Research Office under 
contract/grant number W911NF-25-2-0060 and by the U.S. Department of Energy, Office of Science, Office of Basic Energy, Materials Sciences and Engineering Division under Award Number DE‐SC0017717. M.F. acknowledges financial support from EPSRC project ID EP/X035158/1; EPSRC/STFC/NSERC under UK-Canada Quantum for Science Research Collaborations OPP640-APP47830 and from Dstl under DASA Contract No: ACC2212159 for CY2025 – Cycle 2 – Defence Rapid Impact
\bigskip

\bibliography{refs}

\end{document}


\begin{center}
{\large\textbf{Supplementary Information}}\\[4pt]
{\normalsize Nonlinearity Reversal in Epsilon-Near-Zero Indium Tin
Oxide Driven by Few-Cycle Light Pulse}
\end{center}

\vspace{8mm}

\section{Pump pulse characterisation}

\begin{figure}[h]
\centering
\includegraphics[width=0.85\textwidth]{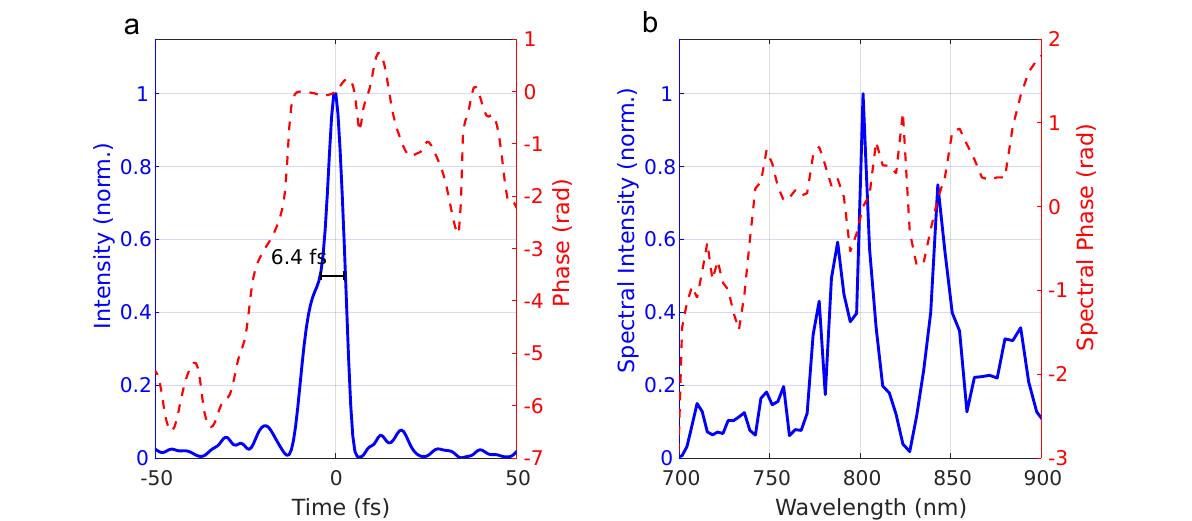}
\caption{\textbf{Retrieved pump pulse profiles.}
(a) Temporal intensity (blue) and phase (red dashed) showing a 6.4\,fs FWHM pulse.
(b) Spectral intensity and spectral phase across the 700--900\,nm bandwidth.}
\label{fig:SI1}
\end{figure}

\begin{figure}[h]
\centering
\includegraphics[width=0.85\textwidth]{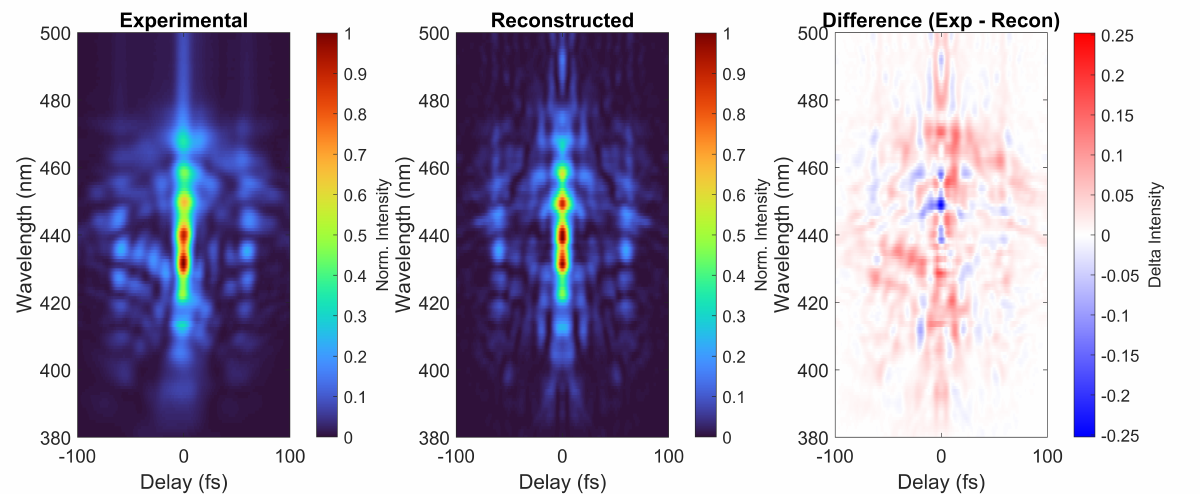}
\caption{\textbf{SHG-FROG autocorrelation traces.}
Experimental trace (left), reconstructed trace (centre), and difference map
(right). The small residuals confirm a reliable retrieval.}
\label{fig:SI2}
\end{figure}

The pump pulse was characterised using second-harmonic generation
frequency-resolved optical gating (SHG-FROG). The retrieved temporal and spectral
profiles are shown in Fig.~\ref{fig:SI1}. The FROG algorithm converges to a pulse
duration of 6.4\,fs FWHM. The experimental and reconstructed FROG autocorrelation traces are shown
in Fig.~\ref{fig:SI2}, along with the difference map. The small and structureless
residuals confirm the quality of the retrieval.

\clearpage

\section{Transmitted probe spectra}

\begin{figure}[h]
\centering
\includegraphics[width=0.95\textwidth]{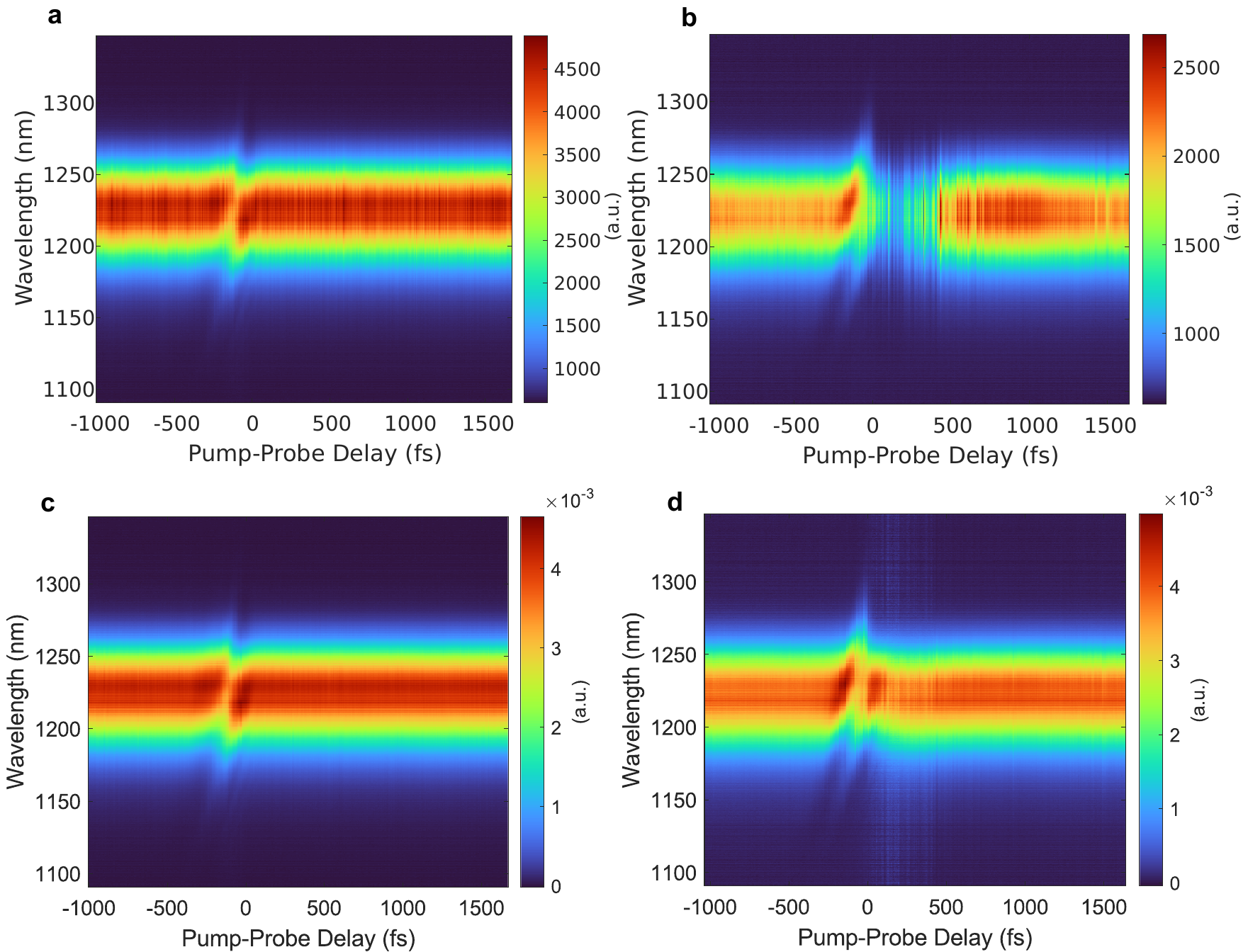}
\caption{\textbf{Transmitted probe spectra as a function of pump-probe delay.}
(a,b) Raw spectral intensity for bare glass and ITO respectively.
(c,d) Normalised spectra for bare glass and ITO. In bare glass only a transient
spectral shift is observed at time zero, while ITO shows a persistent intensity
modulation confirming a genuine nonlinear optical response.}
\label{fig:SI3}
\end{figure}

Figure~\ref{fig:SI3} shows the transmitted probe spectrum as a function of
pump-probe delay for bare glass (left column) and ITO (right column) under the highest intensity pump pulse ($10$ $TW/cm^2$). The top row
shows raw spectral intensity and the bottom row shows the spectra normalised such
that the total intensity at each delay is set to unity. In bare glass, cross-phase
modulation produces a transient spectral shift around time zero but no persistent
intensity modulation. In ITO, both a spectral shift and a sustained intensity
modulation are observed, confirming that the signal measured in the main text
reflects genuine population dynamics that happens after pump-probe overlap in the ITO film rather than a coherent
interaction artefact. 

\clearpage

\section{Longer Pulse Simulation}

\begin{figure}[h]
\centering
\includegraphics[width=0.85\textwidth]{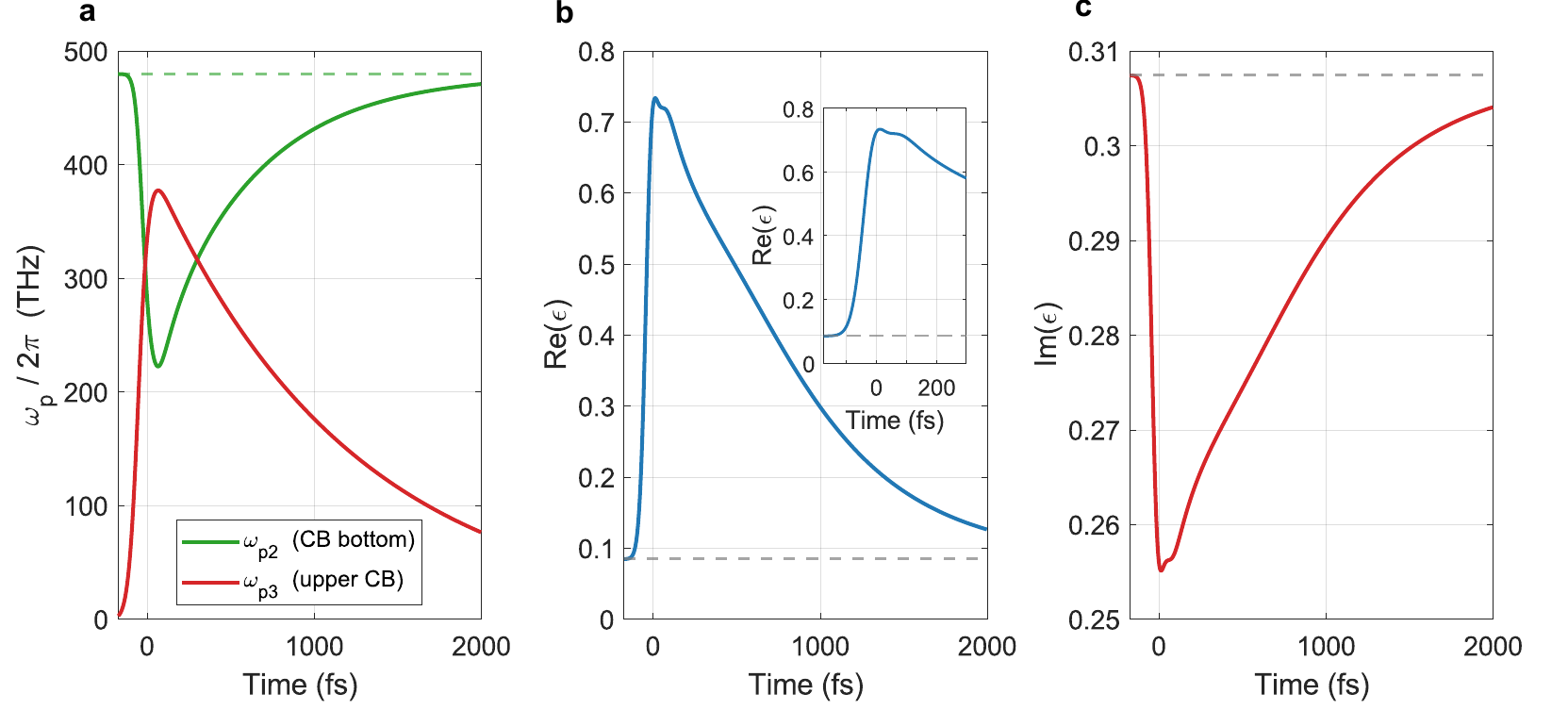}
\caption{\textbf{Modelled permittivity dynamics under 100\,fs excitation.}
(a) Time-dependent plasma frequencies $\omega_\mathrm{p2}$ and $\omega_\mathrm{p3}$.
(b) Real part of the permittivity Re($\varepsilon$) and (c) is the imaginary part Im($\varepsilon$) of the permittivity. Under these conditions
Re($\varepsilon$) remains positive throughout and decays monotonically, consistent
with a purely intraband response.}
\label{fig:SI4}
\end{figure}

\begin{figure}[h]
\centering
\includegraphics[width=0.95\textwidth]{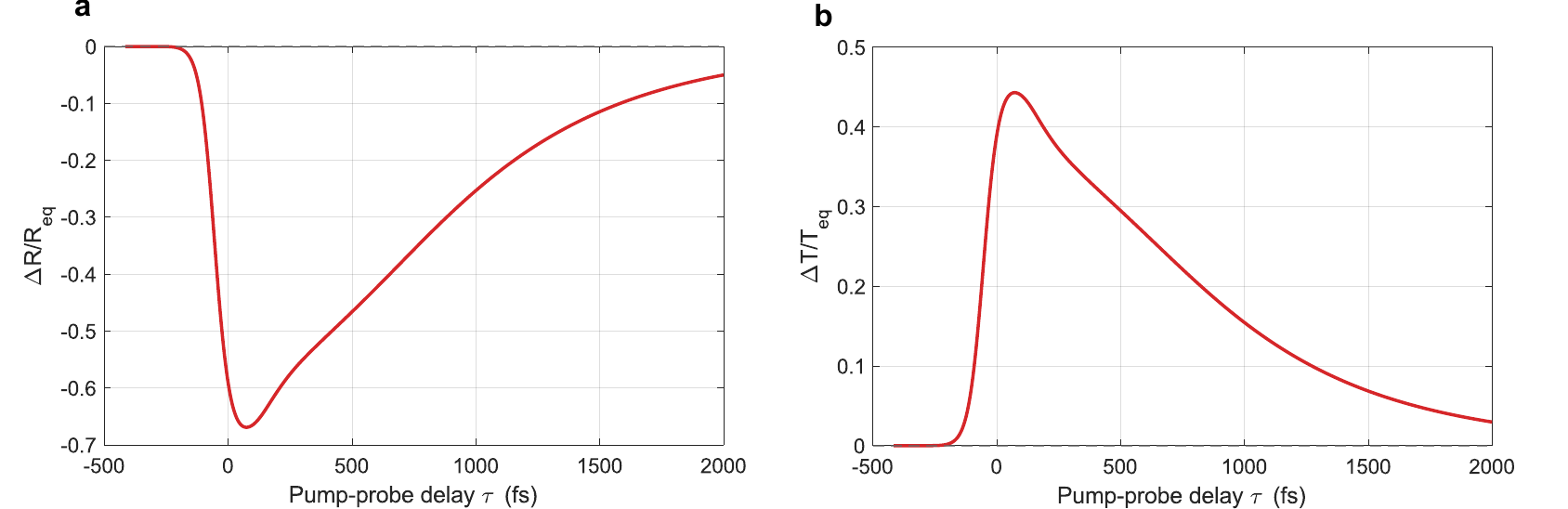}
\caption{\textbf{Transient optical response under 100\,fs excitation at equivalent
fluence.} (a) $\Delta T/T$ and (b) $\Delta R/R$ for a 100\,fs, 1.1\,TW/cm$^{2}$
pump pulse, matched in fluence to the highest intensity measurement in the main
text. No sign reversal or secondary feature is observed.}
\label{fig:SI5}
\end{figure}

To confirm that the sign reversal observed in the main text is driven by peak
intensity rather than pulse fluence, we repeated the simulations with retrieved model parameters for
a 100\,fs FWHM pump pulse at 1.1\,TW/cm$^{2}$, matched in fluence to the highest
intensity case in the main text. The transient reflectance and transmittance,
shown in Fig.~\ref{fig:SI4}, remain monotonic throughout, with no secondary
feature or sign reversal. The corresponding time-dependent plasma frequencies and
complex permittivity from the rate equation model are shown in Fig.~\ref{fig:SI5},
confirming that Re($\varepsilon$) remains positive and decays monotonically under
these conditions, never crossing the ENZ point. These results demonstrate that
peak intensity, not fluence, is the critical parameter governing the onset of the
competing nonlinear channel.

